\documentclass{ws-ijmpd}
\usepackage{epsfig}
\usepackage{epstopdf}
\usepackage{color}

\begin{document}
	
\title{Wormhole solutions in $f(R)$ Gravity}

\author{B. Mishra}
\address{Department of Mathematics, Birla Institute of Technology and Science-Pilani, Hyderabad Campus, Hyderabad 500078, India\\ bivu@hyderabad.bits-pilani.ac.in}

\author{A.S. Agrawal}
\address{Department of Mathematics, Birla Institute of Technology and Science-Pilani, Hyderabad Campus, Hyderabad 500078, India\\ agrawalamar61@gmail.com}

\author{S.K. Tripathy}
\address{Department of Physics, Indira Gandhi Institute of Technology, Sarang, Dhenkanal, Odisha 759146, India \\tripathy\_sunil@rediffmail.com}

\author{Saibal Ray}
\address{Department of Physics, Government College of
	Engineering and Ceramic Technology, Kolkata 700 010, West Bengal, India\\ saibal@associates.iucaa.in}

\maketitle	

\abstract{In this work, we have studied the traversable wormholes geometry in $f(R)$ theory gravity, where $R$ be the Ricci scalar. The wormhole solution for some assumed $f(R)$ functions have been presented. The assumption of $f(R)$ is based on the fact that its behaviour changed with an assumed parameter $\alpha$ rather than the deceleration parameter. Three models are presented based on the physically motivated shape function and their behaviours are studied. }


\section{Introduction}
The pioneering works on static traversable wormholes generated considerable amount of research in the last couple of decades~\cite{Ellis73,Bronnikov73,Morris88,Hochberg1993,Visser1989,Visser1987,Kim2001,Dadhich2002,Kuhfitig2003,Arellano2006,Arellano2006a,Garattini2019,Tripathy2021}. In principle, wormholes connect two distant parts of the same universe or can also connect two universes. The wormholes are hypothetical tunnels and look like a tube and flat on both sides asymptotically. Most significantly, Einstein and Rosen~\cite{Einstein35} have presented the mathematical model that connects two asymptotically flat universe through a bridge which is popularly known as the Einstein-Rosen bridge. Einstein and Rosen~\cite{Einstein35} observed that an exotic matter may be present to produce anti-gravity effect that may provide some sort of stability, otherwise due to gravity, the wormhole  may collapse. In view of this, exotic matter violating the null energy condition (NEC) should exist around the wormhole throat~\cite{Visser03,Fewster05,Kuhfittig06,Jamil10}.

The wormhole solution by Morris and Thorne~\cite{Morris88} was based on the standard general relativity (GR) and presented significant improvement on the earlier wormhole geometry such as Wheeler wormholes~\cite{Garfinkle1991,Visser1991}, Kerr wormholes~\cite{Bueno2018,Jusufi2019} and Schwarzschild wormholes~\cite{Collas2012,Cataldo2017}.  Lemos et al.~\cite{Lemos03} have presented a systematic review of wormhole solutions and  matched the interior solution to the unique exterior vacuum solution using Einstein's equations whereas Hayward~\cite{Hayward09} reviewed the dynamic processes that involves wormholes and suggested the wormhole thermodynamics. Sarbach and Zannias~\cite{Sarbach10} have claimed the stability of their static wormhole solutions with redial magnetic field and exotic dust with respect to radial fluctuations. Bronnikov et al.~\cite{Bronnikov13} have suggested some symmetric wormhole model whose metric coincides with Ellis wormhole~\cite{Ellis73}. Shinkai and Hayward have studied the stability of the Morris-Thorne traversable wormholes in the presence of a massless ghost Klein-Gordon field~ \cite{Shinkai2002}.  Bejarano et al.~\cite{Bejarano17} have presented two classes of traversable wormhole space-times with a single matter source defined in the action. Moreover, the space-times presented are exact solutions of some extension of GR. Cremona et al.~\cite{Cremona19} have tested and compared the linear instability of Morris-Thorne~\cite{Morris88}, Ellis~\cite{Ellis73}. Bronnikov et al.~\cite{Bronnikov13} have studied a static spherically symmetric wormhole with perfect fluid having negative energy density and a source free radial electric or magnetic field. The wormholes studied by them are obtained to be stable against spherically symmetric and axial perturbations.  A few interesting models are available in literature in connection to galactic wormholes~\cite{kuhfittig2014,rahaman2014a,rahaman2014b,rahaman2016a,rahaman2016b,chakraborty2021} based on mainly rotation curves.

The initial motivation for $f(R)$ theory of gravity was based on the inflationary scenario~\cite{Starobinsky80}. In fact, Carroll et al.~\cite{Carroll04} have explained the late time cosmic acceleration in the context of $f(R)$ gravity. Subsequently $f(R)$ theory of gravity has been incorporated in different context. Several researchers have studied various viable cosmological models in $f(R)$ gravity~\cite{Cognola05,Faraoni05,Bergliaffa06,Capozziello06,Amarzguioui06,Santos07,Koivisto07,Amendola07,Tsujikawa08,Ananda08,Carloni08}. Another aspect was to explore the coupling of an arbitrary function of $R$ with the matter Lagrangian density~\cite{Bertolami07,Sotiriou08,Harko08}. Constraining from strong lensing, Yang and Chen~\cite{Yang09} studied $f(R)$ gravity in Palatini formalism. From $f(R)$ gravity viewpoint, Capozziello and Laurentis~\cite{Capozziello12} have given an alternate approach to dark matter problem. A few notable works under $f(R)$ gravity are as follows: Tripathy and Mishra~\cite{Tripathy16} presented a vacuum solution with reconstructed Ricci scalar, Mongwane~\cite{Mongwane17} formulated the characteristic initial value problem for $f(R)$ gravity, Liu et al.~\cite{Liu18} have constrained the $f(R)$ gravity in cosmology, solar system as well as binary pulsar systems, and Lazkoz et al.~\cite{Lazkoz18} reformulated $f(R)$ Lagrangian terms as an explicit functions of the redshift.  It is worth to mention here that,  Bronnikov and Starobinsky \cite{Bronnikov07} proved that wormholes can not be formed in dark energy models governed by scalar tensor theory even in the presence of electric or magnetic fields.

Beato et al.~\cite{Beato16} and Canfora et al.~\cite{Canfora17} have shown that to construct the traversable wormhole in GR, the exotic matter is not mandatory. On a similar note, other scientists, like Harko et al.~\cite{Harko13}, Pavlovic and Sossich~\cite{Pavlovic15}, also have described that the modified theory of gravity such as $f(R)$ gravity can describe the wormhole geometry without invoking any exotic matter. De Benedictis and Horvat~\cite{DeBenedictis12} have shown the existence of wormhole throat in modified $f(R)$ gravity and studied its properties whereas Sharif and Zahra~\cite{Sharif13}  explored the wormhole solutions for isotropic, anisotropic fluids and barotropic equation of state with the radial pressure. It is noted that Eiroa and Aguirre~\cite{Eiroa16} have presented thin-shell Lorentzian wormholes in $f(R)$ gravity and analyzed the stability of the model. In $f(R)$ gravity Mazharimousavi and Halilsoy~\cite{Mazharimousavi16} have constructed the  traversable wormholes model that satisfies the energy conditions. Bhattacharya and Chakraborty~\cite{Bhattacharya17} presented the wormhole solution supported with two matter components, such as (i) the homogeneous and isotropic, and (ii) the inhomogeneous and anisotropic in $f(R)$ gravity. Golchin and Mehdizadeh~\cite{Mehdizadeh19} have obtained the exact solutions of traversable wormholes with the non-constant Ricci scalar whereas Restuccia and Tello-Ortiz~\cite{Restuccia20} have given new class of $f(R)$ gravity model and studied the cosmological parameters. In a recent work, Godani and Samanta~\cite{Godani20} have studied the traversable wormhole solutions both in variable and constant redshift functions in $f(R)$ gravity.

The paper is organized as follows: the basic formalism of $f(R)$ gravity and the corresponding field equations are derived in Section 2. In Section 3, we have derived the wormhole solution in $f(R)$ gravity for three different shape functions and the graphical representation are made. In Section 4, we have presented our important results and discussions.

\section{Basic Formalism and Field Equations}
The action for $f(R)$ gravity can be considered as
\begin{equation} \label{eq1}
S=\frac{1}{16 \pi}\int d^4x\sqrt{-g}f(R)+\int d^4x\sqrt{-g}\mathcal{L}_m,
\end{equation}
where $g$ and $\mathcal{L}_m$ respectively denote the determinant of the metric $g_{\mu \nu}$ and the matter Lagrangian density. The natural system of unit has been used: $G=\hbar=c=1$, where $G$, $\hbar$ and $c$ represent the Newtonian gravitational constant, reduced Planck constant and speed of light in vacuum respectively.

By varying the action with respect to $g^{\mu\nu}$ and using the metric approach, we can obtain  
\begin{equation} \label{eq2}
FR_{\mu \nu}-\frac{1}{2}fg_{\mu \nu}-\nabla_{\mu}\nabla_{\nu}F+g_{\mu \nu}\square F=T_{\mu \nu},
\end{equation}
where $F=\frac{df(R)}{dR}$. 

We can consider the contraction of Eq. \eqref{eq2} to obtain the relation
\begin{equation} \label{eq3}
FR-2f+3\square F=T,
\end{equation} 
where $R=g^{\mu\nu}R_{\mu\nu}$ and $T=g^{\mu\nu}T_{\mu\nu}$ are respectively the Ricci scalar and trace of stress energy tensor.

Now, substituting Eq. \eqref{eq3} and rearranging the terms, the Einstein field equations for $f(R)$ gravity can be derived as~\cite{Lobo09}
\begin{equation} \label{eq4}
G_{\mu \nu}=R_{\mu \nu}-\frac{1}{2}Rg_{\mu\nu}=T_{\mu\nu}^{eff},
\end{equation}
where $T_{\mu\nu}^{eff}$ is the effective stress energy tensor which is the combination of stress energy tensor $T_{\mu \nu}^{(c)}$ and $\widehat{T}_{\mu \nu}^{(m)}$ along with the following expanded forms:
\begin{eqnarray} \nonumber
\widehat{T}_{\mu \nu}^{(m)}&=&{T}_{\mu \nu}^{(m)}/F, \\ \nonumber
T_{\mu \nu}^{(c)}&=& \frac{1}{F}\left[\nabla_{\mu}\nabla_{\nu}F-\frac{1}{4}(RF+\square F+T)g_{\mu \nu}\right].
\end{eqnarray}

We are interested to study the static and spherically symmetric traversable wormhole, described as
\begin{equation}\label{eq5}
ds^2=-e^{2\Phi(r)}dt^2+\frac{1}{1-b(r)/r}dr^2+r^2(d\theta^2+sin^2\theta d\phi^2),
\end{equation}
where $\Phi(r)$ and $b(r)$ are two arbitrary functions dependent on the radial coordinate $r$ and respectively known as the redshift function and shape function. The shape function basically describes the geometry of the wormhole under consideration.  The wormhole throat corresponds to a minimum value of the radial coordinate, usually denoted by $b_0$ or $r_0$ in the literature, so that two coordinate patches are now required, each covering the range $[b_0;+1)$, one for the upper universe and another for the lower universe. The redshift function $\Phi(r)$ and the shape function $b(r)$ may actually be different on the upper and lower portions of the wormhole and can be denoted by $\Phi_{\pm}(r)$ and $b_{\pm}(r)$. In this paper, we will assume complete symmetry between the upper and lower parts of the wormhole, thus we consider only one function $\Phi(r)$ and only one function $b(r)$ over the range $[b_0;+1)$~\cite{Lobo07}.

From Eq. \eqref{eq5}, with zero tidal force, i.e. $\Phi=0$, we can derive the Ricci scalar as $R=\frac{2b'(r)}{r^2}$, where the prime denotes a differentiation with respect to $r$. We can define the energy momentum tensor of an anisotropic distribution of matter that describes the exotic matter content of the wormhole in the following form
\begin{equation}\label{eq6}
T_{\mu \nu}=(\rho+ p_t)u_{\mu} u_{\nu}+p_tg_{\mu \nu} +(p_r-p_t)x_{\mu}x_{\nu},
\end{equation} 
where $u^{\mu}$ is the four velocity vector of the fluid that satisfies $u^{\mu} u_{\mu}=-1$ and $x^{\mu}x_{\mu}=1$. Here the symbols $\rho$, $p_r$ and $p_t$ respectively denote the energy density, radial pressure and tangential pressure respectively. So, the trace of the energy momentum tensor can be obtained as $T=-\rho+p_r+2p_t$. 

Varieschi and Ault~\cite{Varieschi16}, after an extensive testing of all possible combinations of the parameters, suggested that the solutions with $\Phi\neq0$ and $\Phi=0$ has not much difference. So here we have, derived the field equations with zero tidal force solution. Then field equations of $f(R)$ gravity \eqref{eq4} for the Morris-Thorne wormhole metric \eqref{eq5} can be derived as
\begin{eqnarray}\label{eq7}
F\frac{b'}{r^2}&=&\rho+H, \\ 
-F\frac{b}{r^3}&=&p_r+r\left(1-\frac{b}{r}\right)\left[F''-F'\frac{b'r-b}{2r(r-b)}\right]-H, \\ \label{eq8}
F\frac{b-b'r}{2r^3}&=&p_t+\left[\left(1-\frac{b}{r}\right)\frac{F'}{r}-H\right], \label{eq9}
\end{eqnarray} 
where the functions $H(r)=\frac{1}{4}(FR+\square F+T)$ and $F=df/dR$. 

Now the whole set of field equations \eqref{eq7}-\eqref{eq9} can be expressed as
\begin{eqnarray} \label{eq10}
\rho&=&F\frac{b'}{r^2},\\
p_r&=&-F\left(\frac{b}{r^3}\right)+F'\left(\frac{b'r-b}{2r^2}\right)-F''\left(\frac{r-b}{r}\right),\\ \label{eq11}
p_t&=&F\left(\frac{b-b'r}{2r^3}\right)+F'\left(\frac{b-r}{r^2}\right).\label{eq12}
\end{eqnarray}

\section{Wormholes Solutions in $f(R)$ Gravity}
We shall study the wormhole solution with the function $f(R)$ expressed as $f(R)=\left[\frac{2}{(\alpha+2)\chi^{\alpha/2}}\right]R^{1+\frac{\alpha}{2}}$, where $\chi=\frac{18}{3+\alpha}-3\left(\frac{k^2+2k+3}{k^2+4k+4}\right)$ with $\alpha$ and $k$ respectively the model parameter and anisotropy parameter~\cite{Tripathy16}. This choice of $f(R)$ is based on the assumption that the behavior of deceleration parameter $q$ does not change with time rather decided by the parameter $\alpha$. With this, we can establish
\begin{eqnarray} \label{eq11}
F&=&\left(\frac{2}{\chi}\right)^{\frac{\alpha}{2}}\left(\frac{b'}{r^2}\right)^{\frac{\alpha}{2}},\\ 
F'&=&F \left(\frac{\alpha}{2}\right)\left(\frac{b''}{b'}-\frac{2}{r}\right),\\ 
F''&=&F \left(\frac{\alpha}{2}\right)\left[\left(\frac{\alpha}{2}-1\right)\left(\frac{b''}{b'}-\frac{2}{r}\right)^2+\left(\frac{b'''}{b'}-\frac{4}{r}\frac{b''}{b'}+\frac{6}{r^2}\right)\right]. 
\end{eqnarray}

To have a reasonable and physically viable models of wormhole, we consider the shape function specifically in the following two forms: (i) $b(r)=r_0+b_1\left(\frac{1}{r}-\frac{1}{r_0}\right)$, (ii) $b(r)=\sqrt{r_0r}$ and (iii) $b(r)=\frac{r_0^2}{r}$. For these choices of the wormhole solutions, we have at the throat, $b(r_0)=r_0$, i.e. the shape function $b(r)$ reduces to the size of the wormhole throat radius $r_0$ at the throat. Other conditions required to be satisfied for the existence of wormhole solution are: (i) $b'(r_0)\leq1$, (ii) $b(r)<r$, for $r>r_0$ and (iii) when $r\rightarrow\infty$, $\frac{b(r)}{r}\rightarrow0$. Obviously, the chosen shape functions satisfy these conditions which are therefore viable for the study of wormholes in the framework of the $f(R)$ gravity.

According to the the gravitational field equations of classical relativity, i.e. GR, the basic research of wormhole geometry is the violation of energy conditions. However, the behavior may vary in the modified theories of gravity. Now, the energy conditions are described by the behavior of the congruence of time-like and space-like curves in Raychaudhuri equation~\cite{Hawking99}. The term $R_{\mu\nu}k^\mu k^\nu$ with $k^\mu$ being the null vector, appears in the space-like curve. When $R_{\mu\nu}k^\mu k^\nu$ is positive, the geodesic congruence remains in the finite value of the geodesic parameters. This resulted the trace of the energy momentum tensor in GR to be positive. However, this may not be possible in the modified theories of gravity since the term $R_{\mu\nu}k^\mu k^\nu$ is not obvious and need to be replaced appropriately in the field equation of modified theories of gravity. In this paper, we have assumed the $f(R)$ theory of gravity, where the things are not straight forward concerning the violation of energy conditions as in GR. So, we have considered three forms of shape function in $f(R)$ gravity to examine the behavior of energy conditions, which may provide some new avenues to study wormhole as a whole. It is to be mentioned here that, the traversable wormhole geometry in the framework of Einstein's GR violates the following conditions, $\rho>0$, NEC-1: $\rho+p_r>0$ and NEC-2: $\rho+p_t>0$. 

\subsection{Case-I}
With the shape function, $b(r)=r_0+b_1\left(\frac{1}{r}-\frac{1}{r_0}\right)$, where $r_0$ is the size of the wormhole throat radius at the throat and $b_1=\frac{\pi^2(8\pi+2\lambda)}{720}$ \cite{Tripathy2021}. The constant $\lambda$ effects the behavior of the shape function $b(r)$. Hence, Eqs. \eqref{eq10}-\eqref{eq12} reduce to
\begin{eqnarray}\label{eq13}
\rho&=& X, \\ 
p_r&=& Y\left[\alpha\frac{b_1^3}{r^{12}}+ A\frac{bb_1^2}{r^{11}}-B\frac{b_1^2}{r^{10}}\right],\\\label{eq14}
p_t&=& Z\left(-\frac{b_1}{r^4}\right)^{\frac{\alpha}{2}}\left[\frac{2\alpha}{r^2}+\left(r_0+\frac{b_1}{r}-\frac{b_1}{r_0}\right)\left(\frac{1-4\alpha}{2r^3}\right)+\frac{b_1}{2r^4}\right]. \label{eq15}
\end{eqnarray}
where $X=\left(\frac{2}{\chi}\right)^{\frac{\alpha}{2}}\left(-\frac{b_1}{r^4}\right)^{\frac{\alpha}{2}+1}$, $Y=\left(\frac{2}{\chi}\right)^{\frac{\alpha}{2}} \left(-\frac{b_1}{r^4}\right)^{\frac{\alpha}{2}-2}$, $Z=\left(\frac{2}{\chi}\right)^{\frac{\alpha}{2}}\left(-\frac{b_1}{r^4}\right)^{\frac{\alpha}{2}}$, $A= (-1+3\alpha+4\alpha^2)$ and $B=2(1\alpha+2\alpha^2)$.

Also, the  energy conditions can be obtained as
\begin{eqnarray}
\rho+p_r&=& Y \left[(\alpha-1)\frac{b_1^3}{r^{12}}+ A\frac{bb_1^2}{r^{11}}- B\frac{b_1^2}{r^{10}}\right], \label{eq16} \\ 
\rho+p_t&=& Y\left[-\frac{b_1^3}{2r^{12}}+(1-4\alpha)\frac{bb_1^2}{2r^{11}}+\frac{2\alpha b_1^2}{r^{10}}\right],\label{eq17}\\ 
p_t-p_r&=& Y \left[ \left(\frac{1-2\alpha}{2}\right)\frac{b_1^3}{r^{12}}+ C\frac{bb_1^2}{r^{11}}+ D \frac{b_1^2}{r^{10}}\right],\label{eq18}\\ 
\frac{p_r}{\rho}&=&\frac{1}{b_1}\left[A br+ B r^2 \right]-\alpha, \label{eq19}\\ 
\rho+p_r+2p_t&=& Y \left[\alpha \frac{b_1^3}{r^{12}}+ E\frac{bb_1^2}{r^{11}}+F\frac{b_1^2}{r^{10}}\right], \label{eq20}
\end{eqnarray}
where $C=(\frac{3}{2}-5\alpha-4\alpha^2)$, $D=4(\alpha+\alpha^2)$, $E=(-\alpha+4\alpha^2)$ and $F= 2(\alpha-2\alpha^2)$.

\begin{figure}[tbph]
\minipage{0.50\textwidth}
\centering
\includegraphics[width=\textwidth]{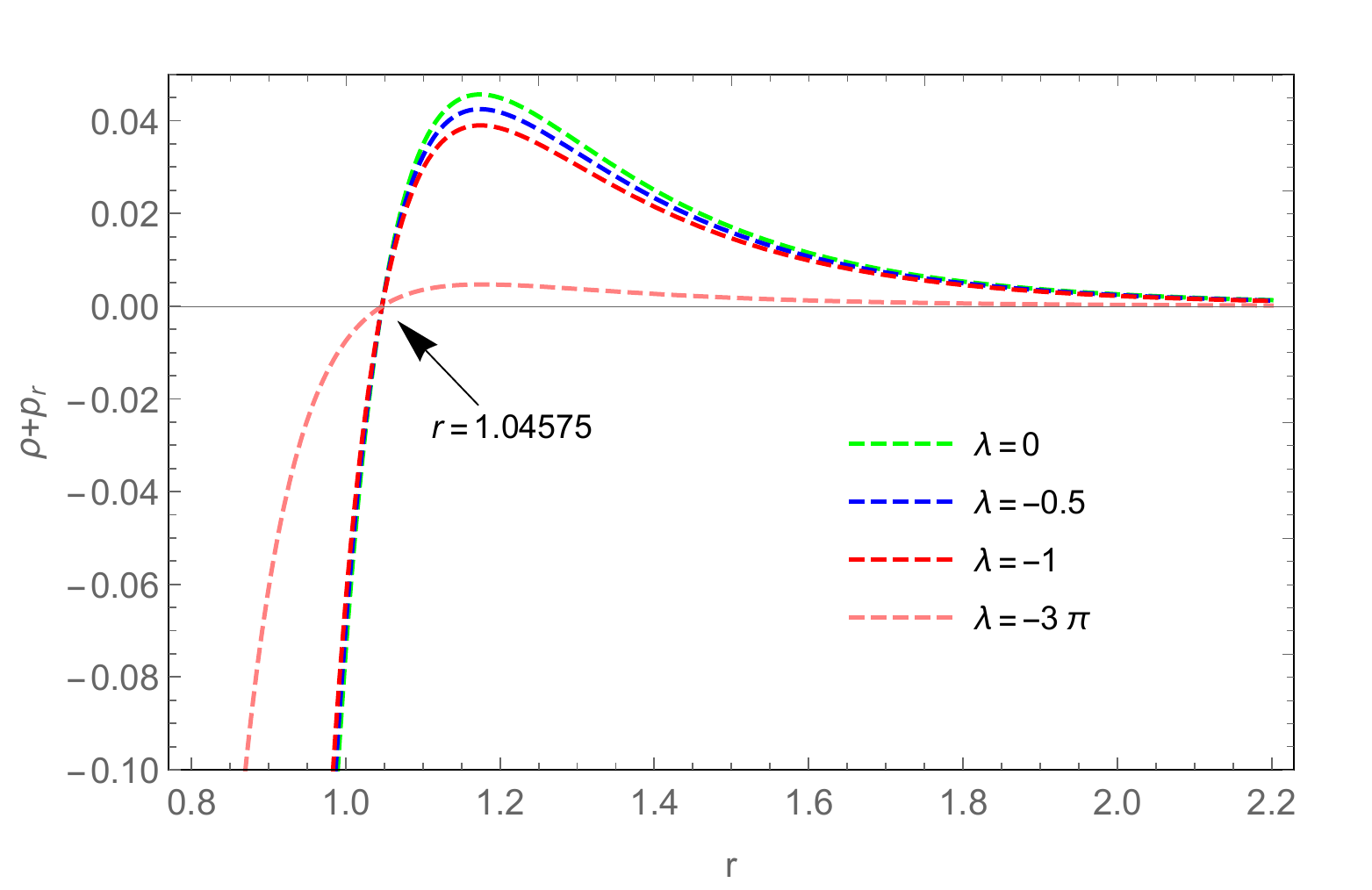}
\caption{$\rho+p_r$ vs $r$~~[Eq. \eqref{eq16}]}
\endminipage\hfill
\minipage{0.50\textwidth}
\includegraphics[width=\textwidth]{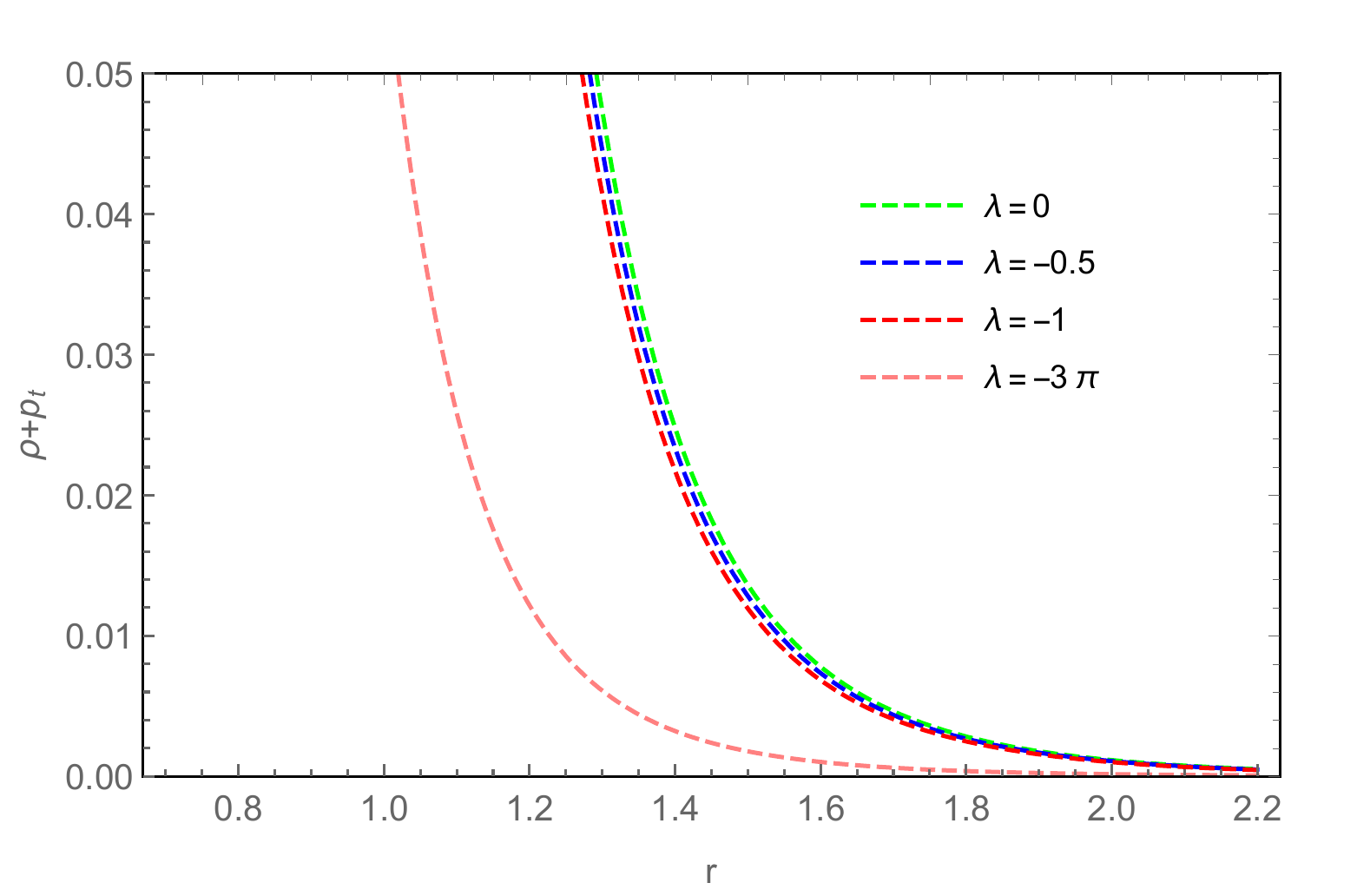} 
\caption{$\rho+p_t$ vs $r$~~[Eq. \eqref{eq17}]}
\endminipage
\end{figure}

The graphical representations of the shape function and energy conditions are exhibited with the representative values of the constant $\lambda=0,-0.5,-1,-3\pi$. The reason behind choosing the representative values is coming out from the conditions imposed on the shape function $b'(r_0)\leq1$ implies $\lambda\geq-4\pi$. The value of anisotropy parameter has been chosen as $1.05$ and the model parameter $\alpha=3$. It is observed that the shape function decreases with the increase in the value of the radial coordinate in all the representative value of $\lambda$, however it always remains positive. Whenever the value of $\lambda$ chosen close to $-4\pi$, the shape function value comes close to $1$, which one can note from $\lambda=-3\pi$. The energy condition $\rho+p_r$ [Fig. 1], indicates the satisfaction and violation of energy condition in certain range of $r$. It violates the energy condition in the range $(0,1.04575)$  and after that it satisfies. Moreover, when the radial value is very high, the shape function vanishes. At the same time, the energy condition $\rho+p_t$ satisfies entirely and vanishes with the increase in the value of $r$ [Fig. 2]. 

The difference in the tangential and radial pressure remains in the positive domain and after $r=1.04575$ [Fig. 3]. It is to note here that the transition is happening at the same point where the energy condition with radial pressure changed its behavior. The difference of the pressure becomes null when $r \geq 2.5$. This shows that the equal amount of pressure observed both in the radial and tangential direction from $r \geq 2.5$. The graphical behavior of the energy condition $\rho+p_r+2p_t$ [Fig. 4] completely satisfies the energy condition and vanishes with the increase in the value of $r$. However, the decrement of this quantity is faster for low values of $\lambda$.

\begin{figure}[tbph]
\minipage{0.50\textwidth}
\centering
\includegraphics[width=\textwidth]{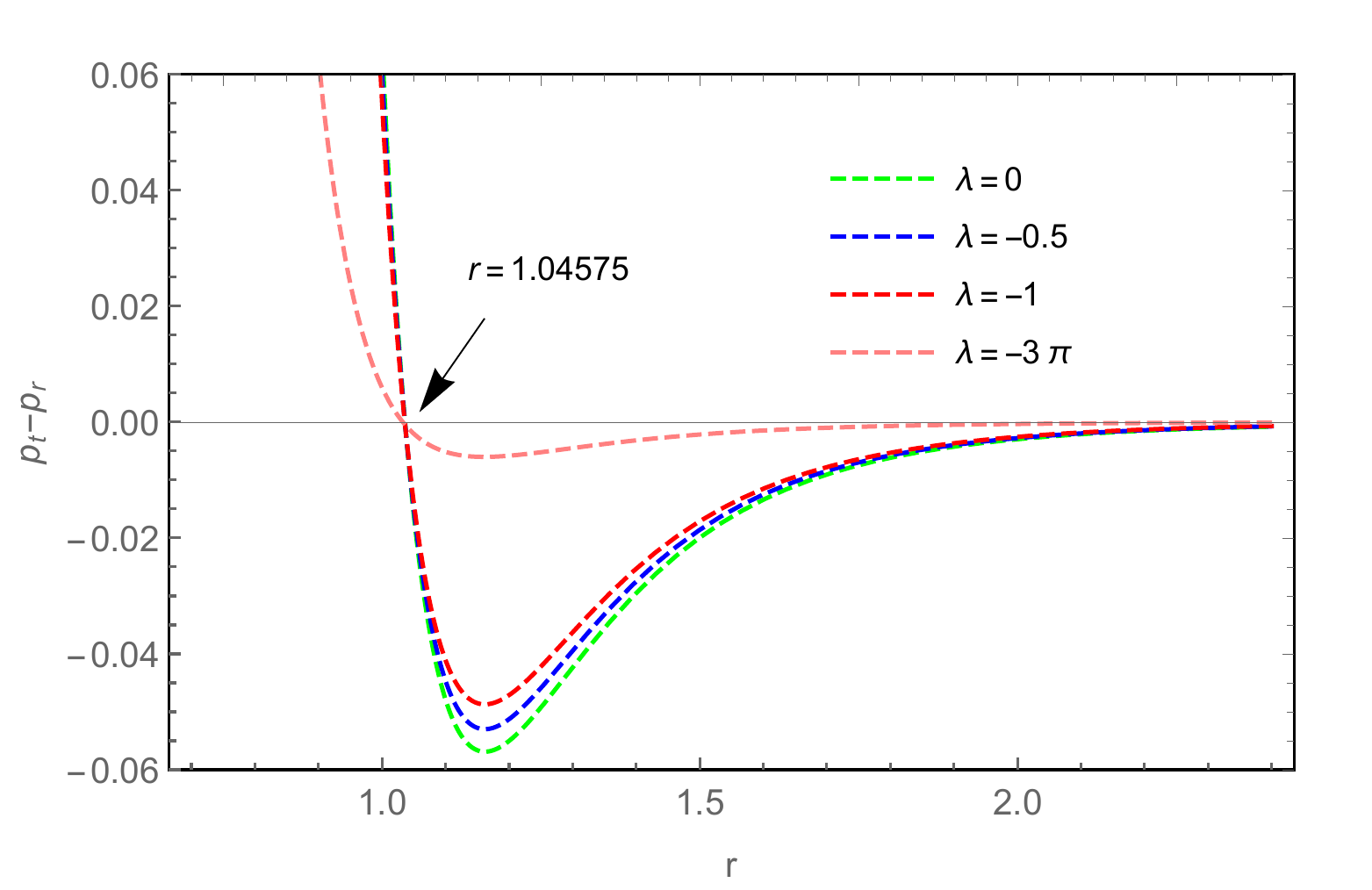}
\caption{$p_t-p_r$ vs $r$~~[Eq. \eqref{eq18}]}
\endminipage\hfill
\minipage{0.50\textwidth}
\includegraphics[width=\textwidth]{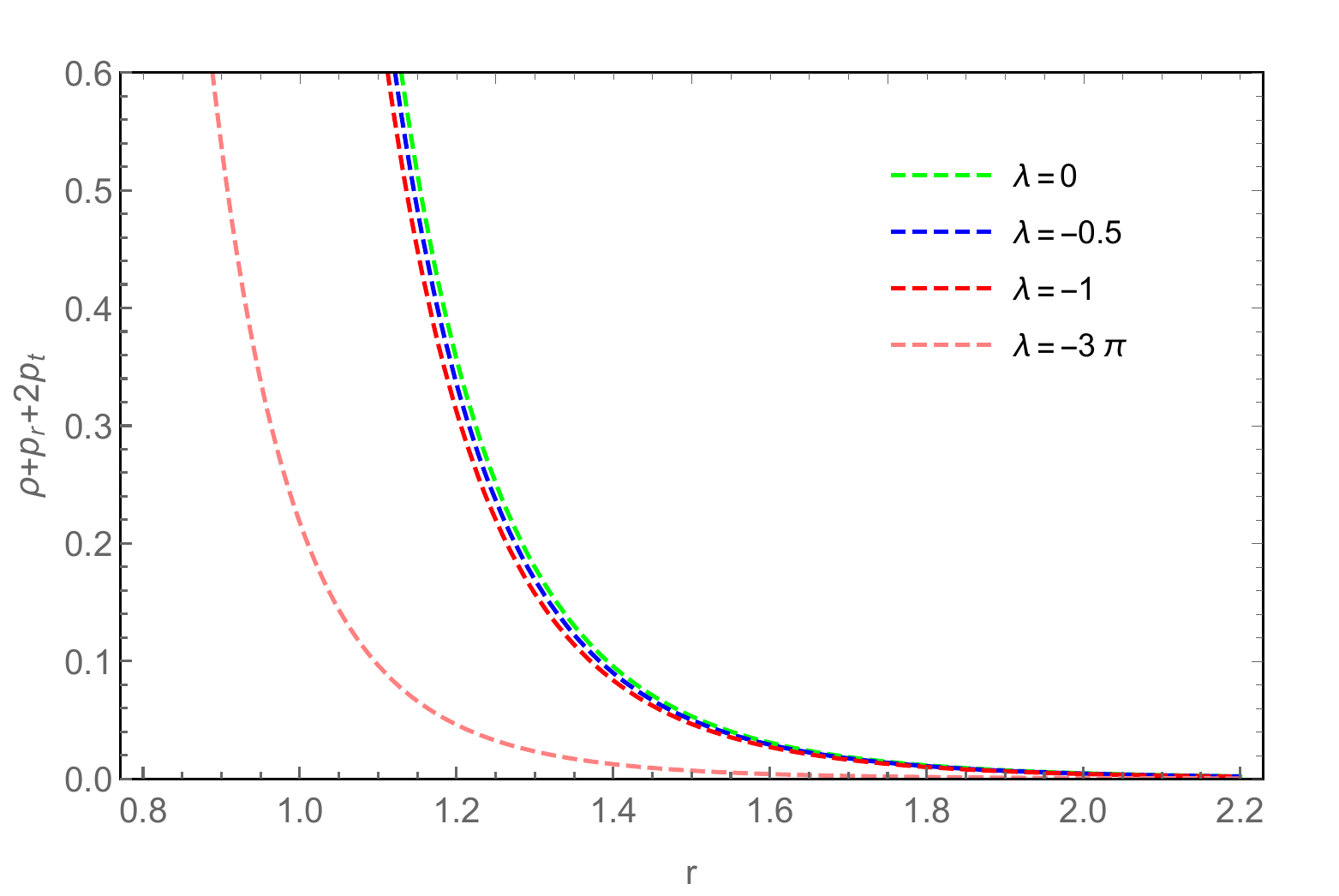} 
\caption{$\rho+p_r+2p_t$ vs $r$~~[Eq. \eqref{eq20}]}
\endminipage
\end{figure}

\subsection{Case-II}
We consider the shape function as $b(r)=\sqrt{r_0r}$ for which the energy density $\rho$, radial component ($p_r$) and tangential component ($p_t$) of the pressure can be derived as
\begin{eqnarray}\label{eq21}
\rho&=&F(r,\chi)\frac{r_0^{\frac{1}{2}}}{2r^{\frac{5}{2}}},\\
p_r&=&F(r,\chi)\left[\left(-1+\frac{25\alpha}{16}+\frac{25\alpha^2}{16}\right)\frac{r_0^{\frac{1}{2}}}{r^{\frac{5}{2}}}-\left(\alpha+\frac{5\alpha^2}{4}\right)\frac{5}{4r^2}\right], \\ \label{eq22}
p_t&=&F(r,\chi)\left[\left(\frac{1-5\alpha}{4}\right)\frac{r_0^{\frac{1}{2}}}{r^{\frac{5}{2}}}+\frac{5\alpha}{4r^2}\right], \label{eq23}
\end{eqnarray}
where $F(r,\chi)=\left[\left(\frac{2}{\chi}\right) \left(\frac{r_0^{\frac{1}{2}}}{2r^{\frac{5}{2}}}\right)\right]^{\frac{\alpha}{2}}$.

Now, the energy conditions can be obtained as
\begin{equation}
\rho+p_r=F(r,\chi)\left[\left(-\frac{1}{2}+\frac{25\alpha}{16}+\frac{25\alpha^2}{16}\right)\frac{r_0^{\frac{1}{2}}}{r^{\frac{5}{2}}}-\left(\alpha+\frac{5\alpha^2}{4}\right)\frac{5}{4r^2}\right], \label{eq24}
\end{equation}

\begin{equation}
\rho+p_t=F(r,\chi)\left[\left(\frac{3-5\alpha)}{4}\right)\frac{r_0^{\frac{1}{2}}}{r^{\frac{5}{2}}}+\frac{5\alpha}{4r^2}\right], \label{eq25}
\end{equation}

\begin{equation}
p_t-p_r=F(r,\chi)\left[\left(\frac{5}{4}-\frac{45\alpha}{16}-\frac{25\alpha^2}{16}\right)\frac{r_0^{\frac{1}{2}}}{r^{\frac{5}{2}}}+\left(2\alpha+\frac{5\alpha^2}{4}\right)\frac{5}{4r^2}\right], \label{eq26}
\end{equation}

\begin{equation}
\frac{p_r}{\rho}= -\left(\frac{25\alpha^2}{8}+\frac{5\alpha}{2}\right)\frac{r^{\frac{1}{2}}}{r_0^{\frac{1}{2}}}+2\left(-1+\frac{25\alpha}{16}+\frac{25\alpha^2}{16}\right), \label{eq27}
\end{equation}

\begin{equation}
\rho+p_r+2p_t=F\left[\left(-\frac{15\alpha}{16}+\frac{25\alpha^2}{16}\right)\frac{r_0^{\frac{1}{2}}}{r^{\frac{5}{2}}}+\left(1-\frac{5\alpha}{4}\right)\frac{5\alpha}{4r^2}\right]. \label{eq28}
\end{equation}

The graphical representations of the null energy conditions and other parameters for the wormhole model are presented with the representative values of $r_0=1,~1.5,~2$. The anisotropy and model parameter values are respectively $k=0.95$ and $\alpha=6$ where the value of anisotropy parameter chosen close to the isotropic value keeping in mind the anisotropy behavior at the late time of the evolution of universe. It can be observed that the shape function increases from zero initially and indefinitely with the increase in the value of the radial coordinate. The increase in the shape function is more rapid for a higher minimum value of the radial coordinate. The NEC-1 violates the null energy condition [Fig. 5], whereas NEC-2 fails to violate [Fig. 6]. This might be due to the different behavior of the radial as well as the tangential pressure, however, both do vanish while the value increases. As a result, in NEC-1, the shape function value increases and approaches to null value whereas in NEC-2 it behaves in a opposite manner. It is worth to mention here that the matter stress-energy condition satisfies the null energy conditions and the violation may happen in the higher derivative of the curvature term~\cite{Lobo09}.

\begin{figure}[tbph]
\minipage{0.50\textwidth}
\centering
\includegraphics[width=\textwidth]{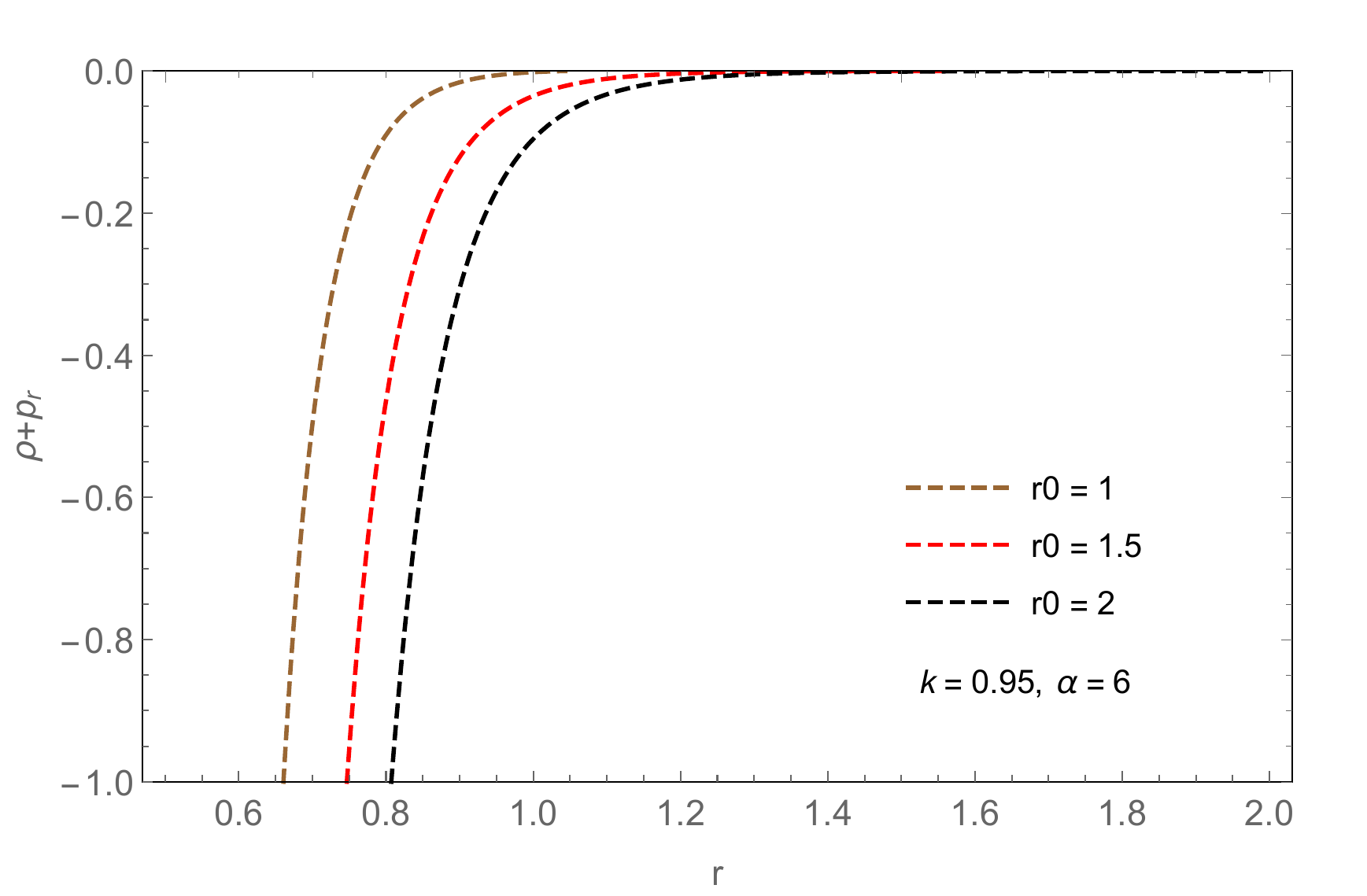}
\caption{$\rho+p_r$ vs $r$~~[Eq. \eqref{eq24}]}
\endminipage\hfill
\minipage{0.50\textwidth}
\includegraphics[width=\textwidth]{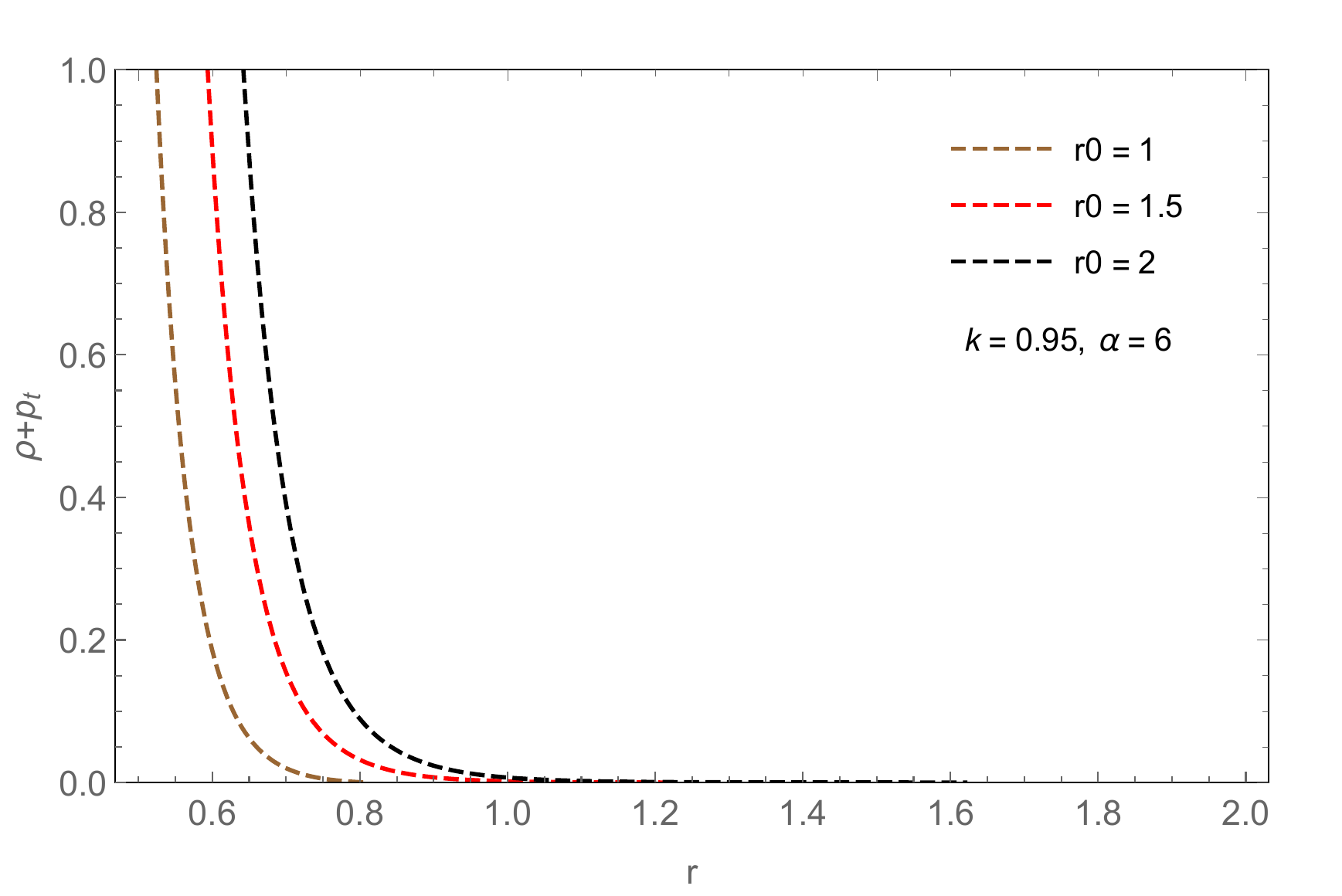} 
\caption{$\rho+p_t$ vs $r$~~[Eq. \eqref{eq25}]}
\endminipage
\end{figure}

The difference between the tangential pressure and the radial pressure decreases steeply from a higher positive value to vanishingly small values with an increase in $r$. It can be inferred that the tangential pressure is dominating the radial pressure [Fig. 7], which may be another reason of the non-violation of NEC-2. At the same time $\rho+p_r+2p_t$ started increasing from a higher negative value and subsequently vanishes [Fig. 8]. We considered the representative values of $r_0$ to see the variation on the behavior of different conditions involved in the study. It has been observed that with the higher value of $r_0$, the curve vanishes later in the radial axis as compare to the lower value.   

\begin{figure}[tbph]
\minipage{0.50\textwidth}
\centering
\includegraphics[width=\textwidth]{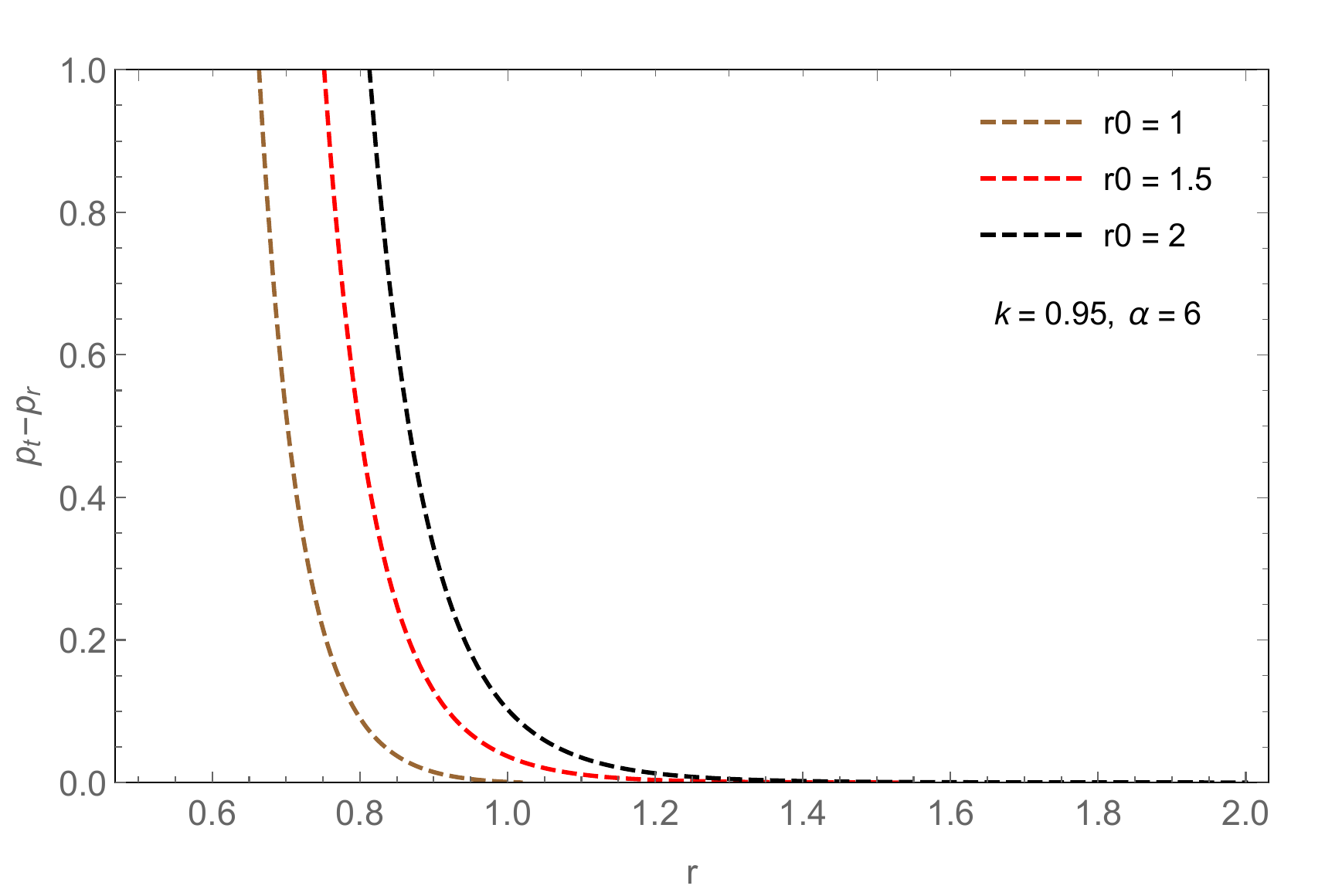}
\caption{$p_t-p_r$ vs $r$~~[Eq. \eqref{eq26}]}
\endminipage\hfill
\minipage{0.50\textwidth}
\includegraphics[width=\textwidth]{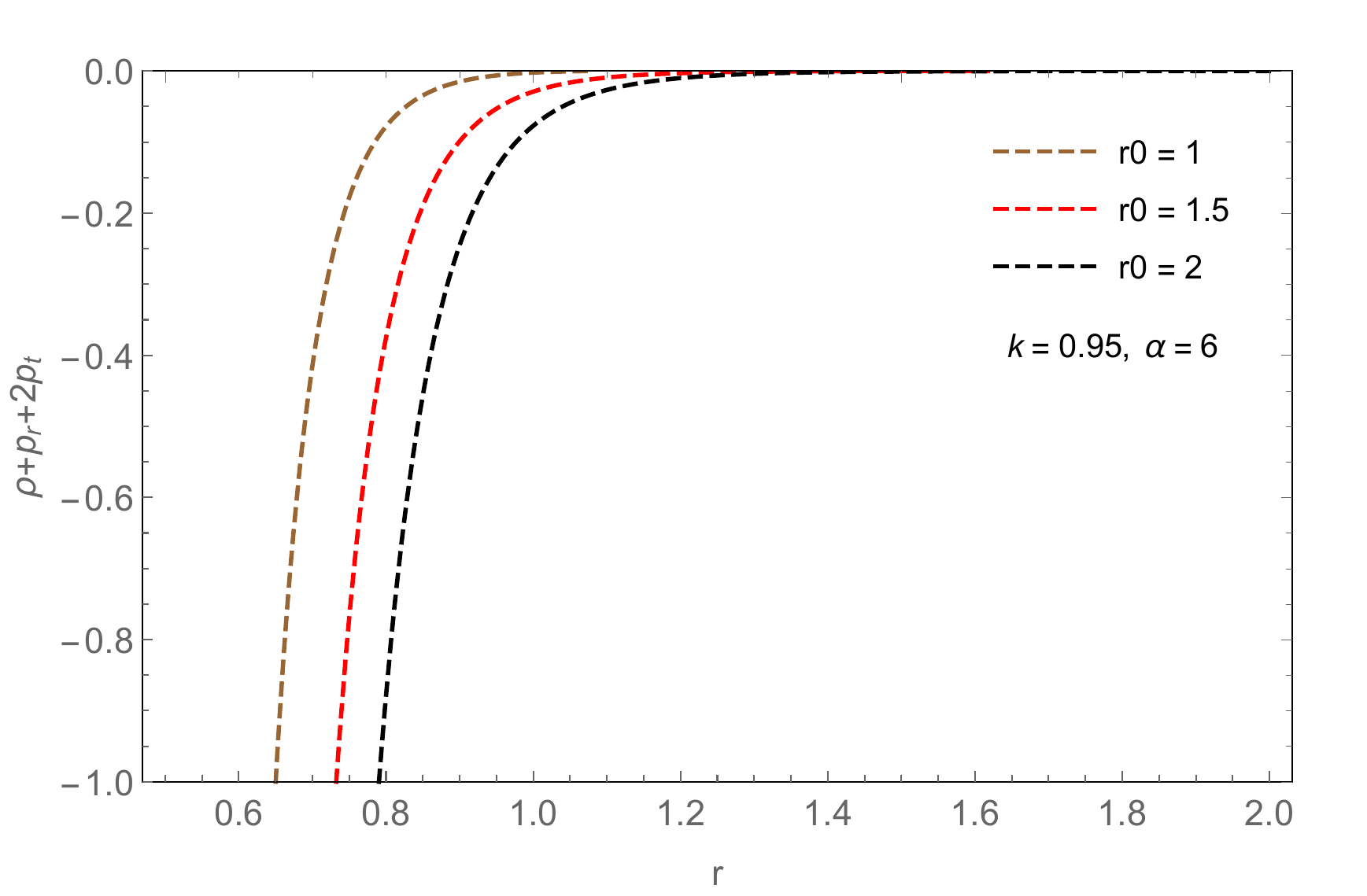} 
\caption{$\rho+p_r+2p_t$ vs $r$~~[Eq. \eqref{eq28}]}
\endminipage
\end{figure}

\subsection{Case-III}
Following Lobo and Oliveira~\cite{Lobo09} we consider the shape function as $b(r)=\frac{r_0^2}{r}$. In this case, the energy density $\rho$, the radial component ($p_r$) and the tangential component ($p_t$) of the pressure can be derived as
\begin{eqnarray}\label{eq29}
\rho&=& G(r,\chi)\left(-\frac{r_0^2}{r^4}\right), \\
p_r&=& G(r,\chi)\left[-\frac{r_0^2}{r^4}+\alpha(\alpha+1)\frac{4r_0^2}{r^4}-\frac{2\alpha}{r^2}(1+2\alpha)\right],\\ \label{eq30}
p_t&=&G(r,\chi) \left[\frac{r_0^2}{r^4}+\left(\frac{\alpha}{2}\right)\left(\frac{4}{r^2}-\frac{4r_0^2}{r^4}\right)\right].\label{eq31}
\end{eqnarray}

Using Eqs. \eqref{eq29}-\eqref{eq31}, the null energy conditions can be derived as
\begin{equation}
\rho+p_r=G(r,\chi)\left[-2\frac{r_0^2}{r^4}+\alpha(\alpha+1)\frac{4r_0^2}{r^4}-\frac{2\alpha}{r^2}(1+2\alpha)\right], \label{eq32}
\end{equation}

\begin{equation}
\rho+p_t=G(r,\chi)\left[\frac{2\alpha}{r}-\frac{2\alpha r_0^2}{r^4}\right], \label{eq33}
\end{equation}

\begin{equation}
p_t-p_r=G(r,\chi) \left[\frac{2r_0^2}{r^4}+\frac{4\alpha}{r^2}-\frac{6\alpha r_0^2}{r^4}+\frac{4\alpha^2}{r^2}-\frac{4\alpha^2r_0^2}{r^4}\right], \label{eq34}
\end{equation}

\begin{equation}
\rho+p_r+2p_t= G(r,\chi)\left[\frac{4\alpha^2r_0^2}{r^4}-\frac{4\alpha^2}{r^2}+\frac{2\alpha}{r^2}\right]. \label{eq36}
\end{equation}

As in the previous case, here we have provided the graphical representation of the null energy conditions with the representative values of $r_0$. The values of the anisotropic and model parameter are assumed respectively as $k=0.95$ and $\alpha=7$ from the physical background of the assumed $f(R)$ function and the shape function. The shape function decreases from higher value and with the increase in the value along the radial axis. The NEC-1 for all the representative values are observed to show the violation of null energy condition [Fig. 9] whereas NEC-2, which is on the tangential pressure direction fails to violate the null energy condition [Fig. 10]. 

\begin{figure}[tbph]
\minipage{0.50\textwidth}
\centering
\includegraphics[width=\textwidth]{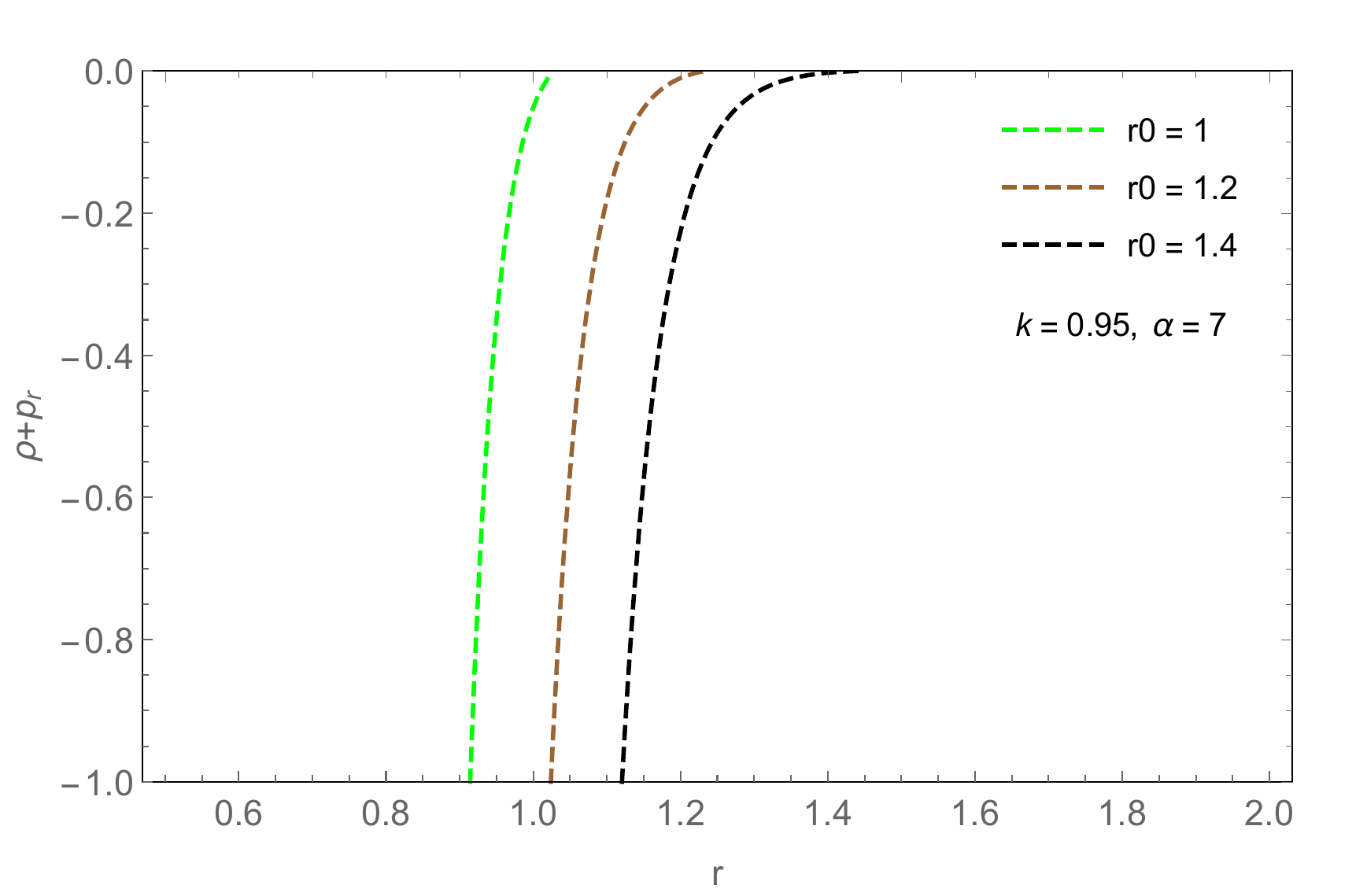}
\caption{$\rho+p_r$ vs $r$ ~~[Eq. \eqref{eq32}]}
\endminipage\hfill
\minipage{0.50\textwidth}
\includegraphics[width=\textwidth]{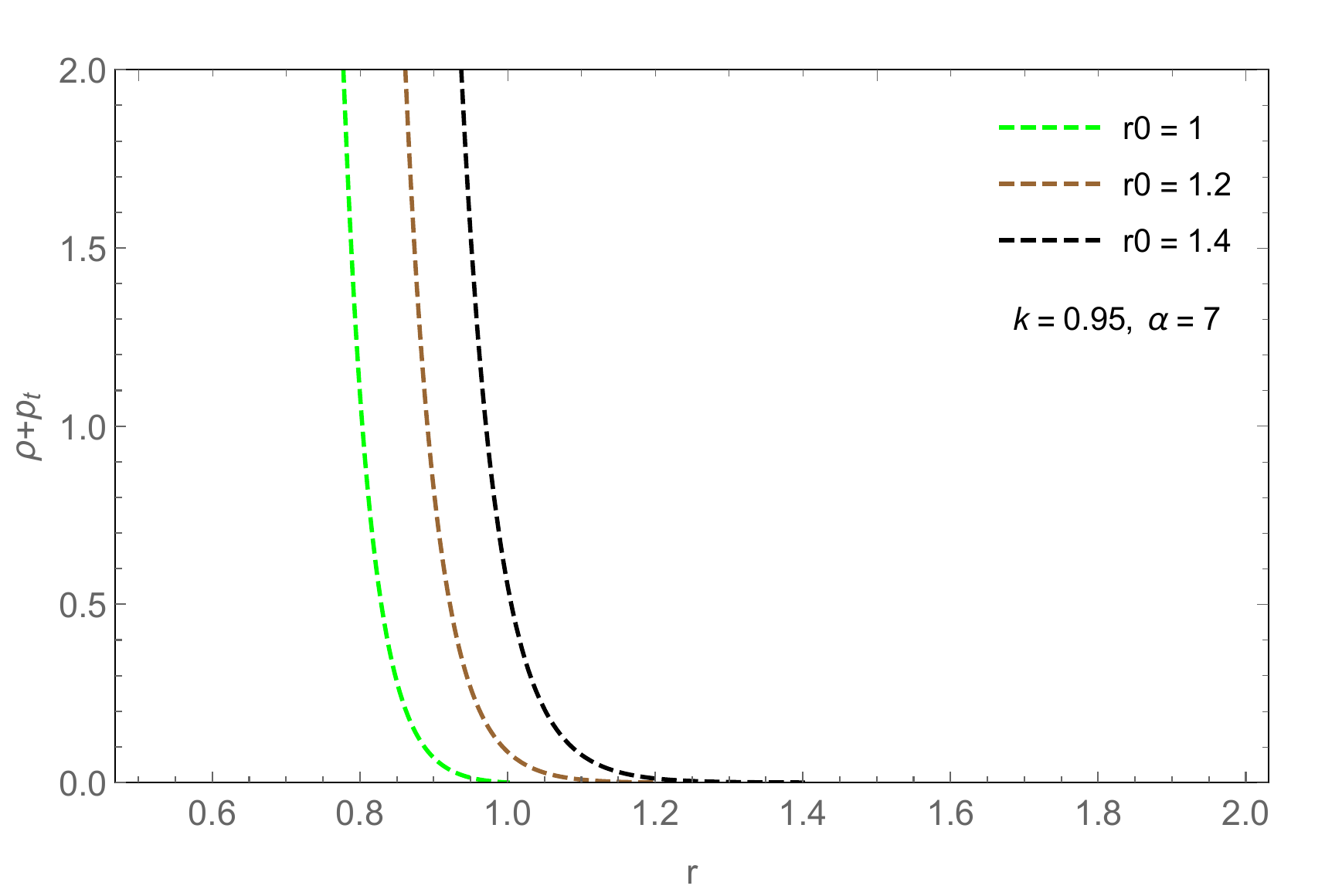} 
\caption{$\rho+p_t$ vs $r$~~[Eq. \eqref{eq33}]}
\endminipage
\end{figure}

The difference between the tangential and radial pressure $p_t-p_r$ decreases from a higher positive value and vanishes with the increase in the value of the radial coordinate [Fig. 11]. At the same time, $\rho+p_r+2p_t$ remains in the negative region and thus violates the energy conditions [Fig. 12]. 

\begin{figure}[tbph]
\minipage{0.50\textwidth}
\centering
\includegraphics[width=\textwidth]{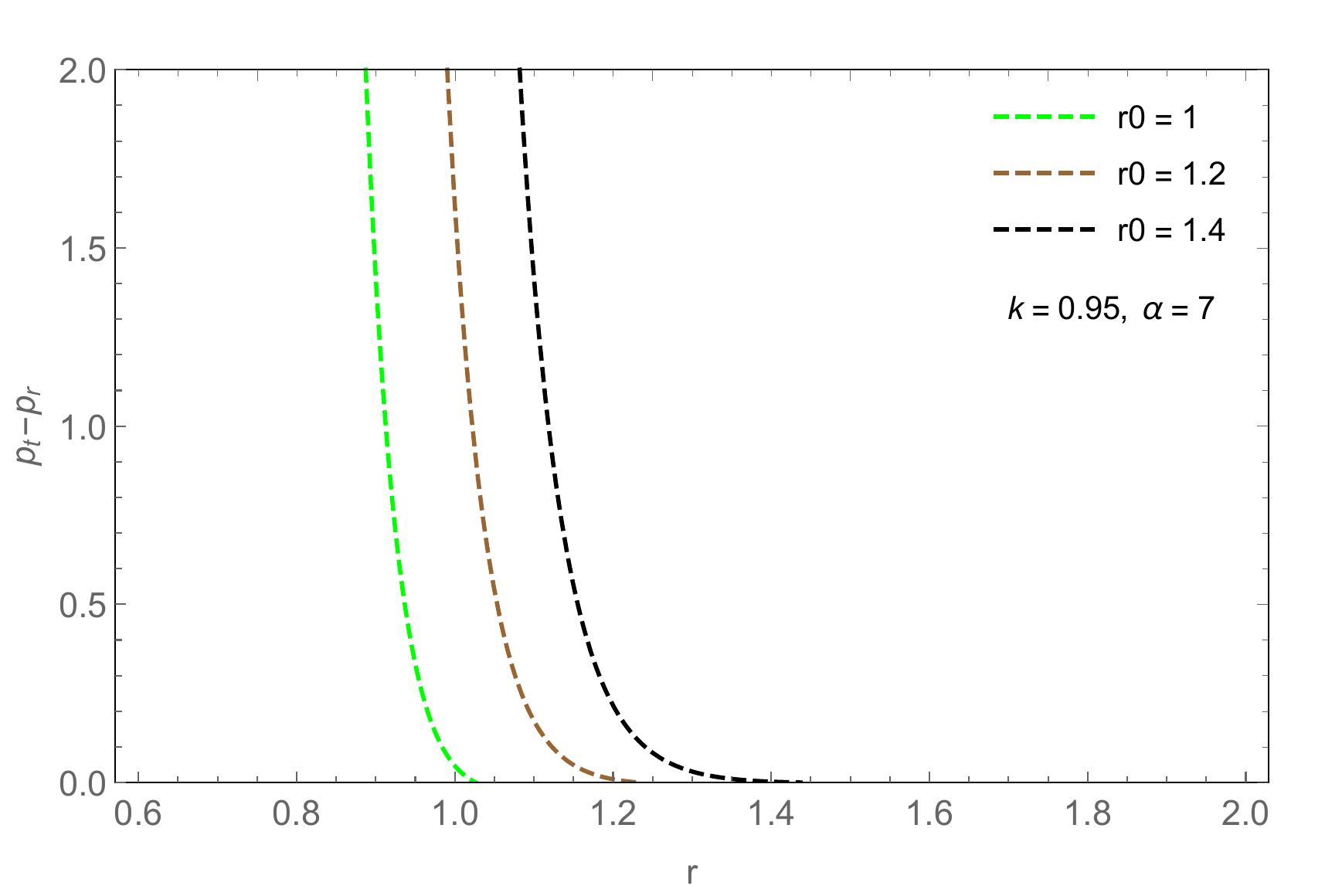}
\caption{$p_t-p_r$ vs $r$ ~~[Eq. \eqref{eq34}]}
\endminipage\hfill
\minipage{0.50\textwidth}
\includegraphics[width=\textwidth]{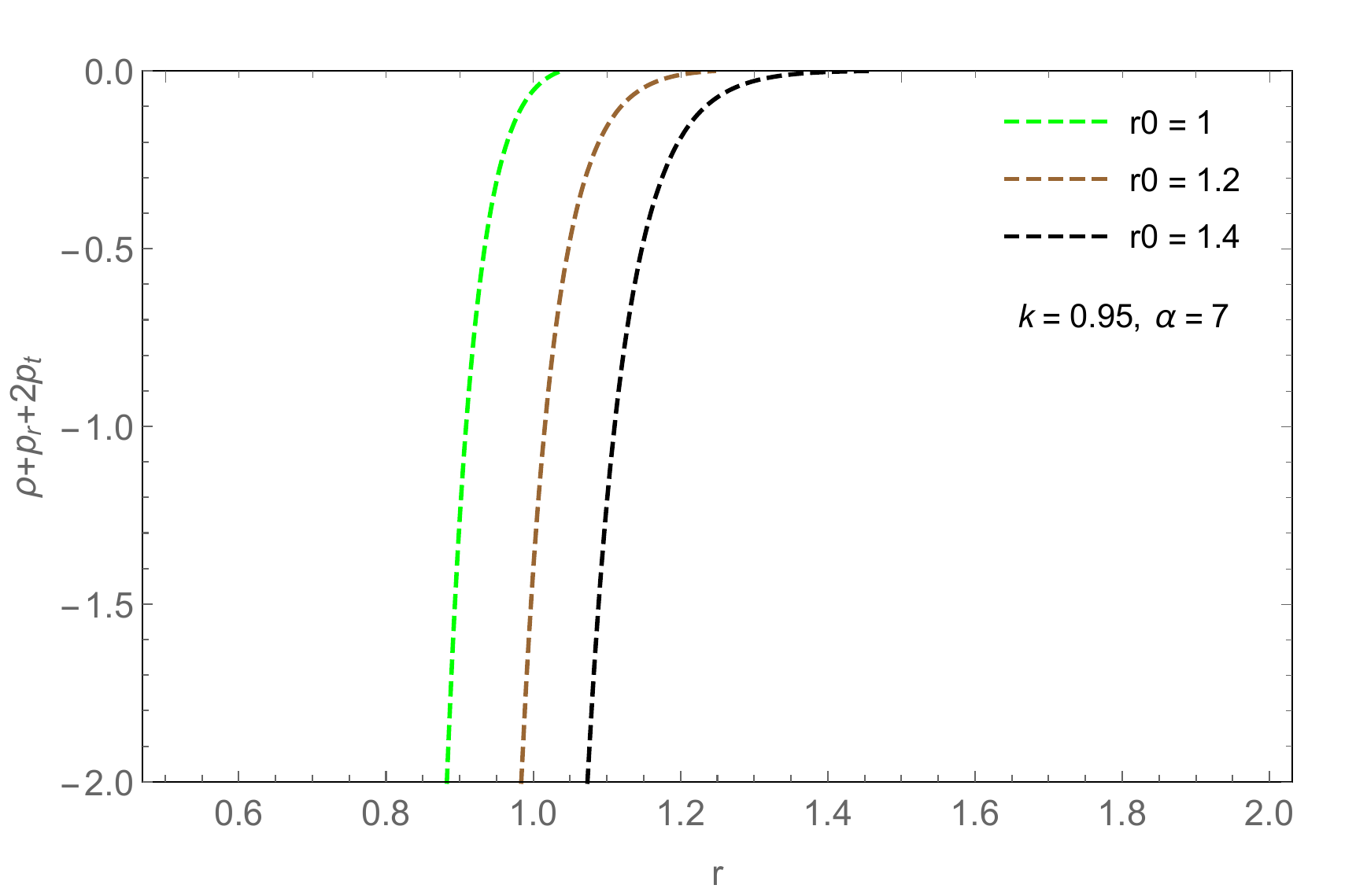} 
\caption{$\rho+p_r+2p_t$ vs $r$ ~~[Eq. \eqref{eq36}]}
\endminipage
\end{figure}

\section{Results and Conclusion} 
As expected the three models presented in the paper show different behavior based on the shape function chosen. As in the literature, the wormhole models should violate the null energy condition when the gravitational field equation is based on Einstein's GR, however it may differ in the case of modified theories of gravity. The same we have observed in our wormhole models. Some of the important results of the models are as follows:\\

(i) Model I shows a transient behavior of the violation of NEC. It violates up to the radial value of $r=1.04575$ and after that it fails to violate. The NEC based on the radial direction looks more convincing compare to the tangential direction for the violation and ultimately satisfying the wormhole physics.\\
(ii) Model II shows the violation of NEC in the energy conditions through the radial pressure whereas for tangential pressure it fails to violate. This may lead to the claim that the behavior may change in $f(R)$ theory of gravity. \\
(iii) The behavior of Model III is similar to Model II though the shape function behaves differently. This indicates the dominance of the functional chosen at the beginning.   \\
(iv) It can be observed that, in all the models $p_r$ is relatively small than $p_t$ which impacts the behavior of null energy conditions.  

Our main motivation in the paper is basically to examine whether wormholes be supported by the $f(R)$ theory of gravity. We have imposed the physically viable functional of $f(R)$ gravity without invoking any exotic matter in the field equations. In the analysis, we have considered some shape functions which have shown the support of wormholes in Einstein's theory of gravitation. As a general conclusion, we would like to mention here that the exact wormhole models can be obtained in the radial equation of the models, however further investigation is required on the tangential equation. 

However, in connection to the present results one may raise the following issue that - is exotic matter indispensable for constructing a traversable wormhole? This query is pertinent as suggestions evolve from the modified theories of gravities that galactic dynamics of massive test particles can be explained without introducing any exotic dark energy~\cite{Carroll2004,Nojiri2007,Nojiri2008,Cognola2008,Cognola2011,Nojiri2011,Deb2018}. This put forward a direct challenge to the results where a exotic component is needed to form a stable wormhole structure. This issue therefore suggest further development of wormhole astrophysics for plausible construction of physically admissible wormholes model. 

Another interesting remark: We have added the comment in the Introduction section for non-zero tidal forces, mentioning that ``From Eq. (5), with zero tidal force, i.e. $\Phi = 0$,…''. It is notable that a non-zero tidal force means an effective pressure will exist due to the non-zero anisotropic factor, i.e. $\Delta \neq 0$ which will be responsible for tidal deformation in the wormhole system and thus may have stability problem. However, this issue of possible calculation of tidal effect via Love numbers can be tackled in a future project~\cite{Das2021}.

\section*{Acknowledgement} BM, SKT and SR thank IUCAA, Pune, India for providing support through the visiting associateship program. ASA acknowledges the financial support provided by University Grants Commission (UGC) through Junior Research Fellowship (File No. 16-9 (June 2017)/2018 (NET/CSIR)), to carry out the research work. The authors are thankful to the authority of the IGIT, Sarang, where the idea was conceived in the National Workshop on Relativity, Cosmology and Astrophysics during January 27-31, 2020.  The authors are thankful to the honourable referees for the constructive comments and suggestions for the improvement of the paper.

\end{document}